\documentclass[aps,preprint,amsmath,amssymb]{revtex4}
\usepackage{graphicx}

\begin{document}

\title
{\Large \bf ATIC/PAMELA anomaly from fermionic decaying Dark
Matter}

\author{ \bf  Chuan-Hung Chen$^{1,2}$\footnote{Email:
physchen@mail.ncku.edu.tw},  Chao-Qiang Geng$^{3}$\footnote{Email:
geng@phys.nthu.edu.tw} and
Dmitry~V.~Zhuridov$^{3}$\footnote{Email:
zhuridov@phys.nthu.edu.tw}
 }

\affiliation{ $^{1}$Department of Physics, National Cheng-Kung
University, Tainan 701, Taiwan \\
$^{2}$National Center for Theoretical Sciences, Hsinchu 300, Taiwan
\\
$^{3}$Department of Physics, National Tsing-Hua University,
Hsinchu 300, Taiwan
 }

\date{\today}

\begin{abstract}
We demonstrate that an economical two Higgs doublet model can
explain the electron and positron excesses in the recent ATIC and
PAMELA experiments by the three body decays of the dark matter
(DM) fermions without requiring the fine turning of the couplings
and degeneracy of masses. We also show that the  mass and lifetime
of the decaying DM particle may not be fixed to be around $1$~TeV
and $10^{26}$~sec, respectively. Moreover, we note that this model
includes a stable dark matter candidate as well.
\end{abstract}

\maketitle

The observed neutrino oscillations~\cite{PDG} and evidence for
dark matter (DM)~\cite{PDG} imply physics beyond the standard
model. In addition, recently reported
PAMELA~\cite{PAMELA}/ATIC~\cite{ATIC} cosmic-ray measurements show
the positron/electron excess above the calculated background for
the energy of order 100 GeV. These data is consistent with the
previous measurements of the high energy electrons and positrons
fluxes in the cosmic ray spectrum by PPB-BETS~\cite{PPB-BETS},
HEAT~\cite{HEAT} and AMS~\cite{AMSCollab}. The, so-called,
PAMELA/ATIC (P/A) anomaly can be explained by either annihilations
or  rare decays of DM particles. However, the possibility of the
DM annihilations requires a boost factor of order $10^2-10^3$ to
make it consistent with the thermally averaged annihilation cross
section obtained from the observed relic density, whereas the
analysis of the DM distribution indicates that the most probable
boost factor should be of order 10~\cite{BoostFactor,Guo_Wu}. On
the other hand, the long enough lifetime of the decaying DM is
achieved by using either some arbitrary small couplings (see
Refs.~\cite{PAMELAanomaly,Guo_Wu,CMSSM,0901.2168} and references
therein) or large scale suppressions associated with
high-dimensional operators at the low energy
in the contexts of supersymmetry~\cite{0812.2075},
technicolor~\cite{Nardi}, hidden gauge boson~\cite{0811.3357} and
hidden fermion~\cite{0812.2374} models.
In particular, DM decays through three-body channels have been discussed
in Refs.~\cite{0812.2075,0812.2374}.
In this letter, we would like to explain the P/A result by three-body decays of
DM fermions in a simple extended
 two-Higgs doublet model.

\begin{figure}[ht]
\centering
\includegraphics*[width=2.8in]{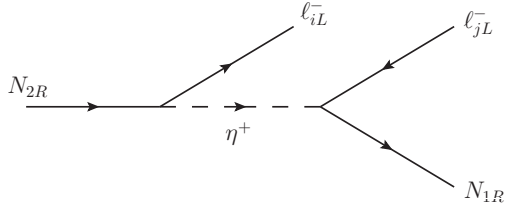}
\caption{Diagram for the DM decay.}
 \label{DMdecay}
\end{figure}

We introduce two neutral leptons $N_k$ with the masses
$M_k$ ($k=1,2$) and second doublet scalar $\eta$ with the mass
$M_\eta$ in addition to the SM particles. We assume that new
particles are odd under a $Z_2$ symmetry. Note that the same
particle content can be used to explain the small neutrino masses
with either stable DM~\cite{Ma} or leptogenesis~\cite{Gu_Sarkar}.
The new Majorana mass term and Yukawa couplings can be written as
\begin{eqnarray}
    M_kN_kN_k + y_{ik}\bar L_i\eta N_{k} + {\rm
    H.c.},
\end{eqnarray}
where $L$ is the lepton doublet
and $i$ and $k$ are the flavor indexes. We consider the masses of
$M_1<M_2<M_\eta$. Hence, $N_1$ is stable and $N_2$ can
only decay to three, as shown in Fig.~\ref{DMdecay}, or more
particles.

The lifetime of $N_2$ is given by
\begin{eqnarray}
    \tau_{N_2}\simeq \frac{1}{\Gamma(N_2\to
    N_1\ell_i^\pm\ell_j^\mp)}=\frac{128(2\pi)^3}{3}\frac{M^4M_2^3}{(M_{21}^2)^4},
\end{eqnarray}
where $M_{21}^2=M_2^2-M_1^2$ is the DM leptons mass splitting and
$M\equiv M_\eta/y$ with $y\equiv|y_{ik}|$. The energy distribution
of electrons/positrons produced in a single three body decay of
$N_2$ can be written as
\begin{eqnarray}
    \frac{dN_{e}}{dE} = \frac{72M_2^3}{(M_{21}^2)^4}\left(M_{21}^2-\frac{16}{9}M_2E\right)E^2.
\end{eqnarray}

We use the same method to calculate the electron/positron flux as
that in
Refs.~\cite{Ibarra_Tran,Chen_Takahashi_Yanagida,Chen_Takahashi}.
The DM component of the primary electron/positron flux is given by
\begin{eqnarray}
    \Phi_{e}^{DM}(E)=\frac{c}{4\pi M_2\tau_{N_2}}\int\limits_0^{M_{21}^2/(2M_2)}dE^\prime G(E,E^\prime)\frac{dN_{e}}{dE^\prime},
\end{eqnarray}
where $E$ is in units of GeV and $c$ is the speed of light. All
the information about astrophysics is encoded in the Green
function $G(E,E^\prime)$, approximately given by
\begin{eqnarray}
    G(E,E^\prime) \simeq \frac{10^{16}}{E^2}\exp[a+b(E^{\delta-1}-E^{\prime\delta-1})]\theta(E^\prime-E) \quad [{\rm cm}^{-3}{\rm s}].
\end{eqnarray}
The coefficients of $a$ and $b$~\cite{PAMELAanomaly} for the spherically
symmetric Navarro, Frenk and White (NFW) density profile of the DM in
our Galaxy~\cite{NFW} and the diffusion parameter $\delta$ are
listed in Table~\ref{T1} for the three propagation models  of M1, MED and
M2~\cite{MED}, respectively, which are
consistent with the observed
Boron-to-Carbon ratio~\cite{BtoC}.
Here, we have neglected the charge-sign
dependent solar modulation~\cite{Blatz} and other astrophysical
uncertainties~\cite{MED}, which could be significant at the energies
below 10~GeV.

\begin{table}[h]
\caption{ Coefficients of the approximate positron Green function
of the NFW halo profile and the diffusion parameter $\delta$ for the
propagation models of M1, MED and M2, respectively.} \label{T1}
\begin{tabular}{|c|c|c|c|}
  \hline
  \ Model\  & $\delta$ & $a$ & $b$ \\
  \hline
  M1 & $\ 0.46\ $ & $\ -0.9809\ $ & $\ -1.1456\ $ \\
  MED & 0.70 & $-1.0203$ & $-1.4493$ \\
  M2 & 0.55 & $-0.9716$ & $-10.012$ \\
  \hline
\end{tabular}
\end{table}

The total electron and positron fluxes are
\begin{eqnarray}
    \Phi_{e^-}=\xi\Phi^{prim}_{e^-}+\Phi^{DM}_{e^-}+\Phi^{sec}_{e^-}
\end{eqnarray}
and
\begin{eqnarray}
    \Phi_{e^+}=\Phi^{DM}_{e^+}+\Phi^{sec}_{e^+},
\end{eqnarray}
respectively, where $\Phi^{prim}_{e^-}$ is a primary astrophysical
component, presumably originated  from supernova remnants,
$\Phi_{e^{-(+)}}^{DM}$ is an exotic primary component from DM
decays, $\Phi_{e^{-(+)}}^{sec}$ is a secondary component from the
spallation of cosmic rays on the interstellar medium, and $\xi$ is
a free parameter about 1 to fit the data when no DM primary source
exists. We take $\xi=0.7$
to insure the flux calculation to be
consistent with the ATIC data.
For the background
fluxes, we will use the parametrizations obtained
in Refs.~\cite{Blatz,Moskalenko}, given by
\begin{eqnarray}\label{Eq_e-}
    \Phi_{e^-}^{prim}(E)&=&\frac{0.16E^{-1.1}}{1+11E^{0.9}+3.2E^{2.15}} \quad
    [{\rm GeV}^{-1}{\rm cm}^{-2}{\rm s}^{-1}{\rm sr}^{-1}],\\
    \Phi_{e^-}^{sec}(E)&=&\frac{0.7E^{0.7}}{1+110E^{1.5}+600E^{2.9}+580E^{4.2}} \quad
    [{\rm GeV}^{-1}{\rm cm}^{-2}{\rm s}^{-1}{\rm sr}^{-1}],\\
    \Phi_{e^+}^{sec}(E)&=&\frac{4.5E^{0.7}}{1+650E^{2.3}+1500E^{4.2}} \quad
    [{\rm GeV}^{-1}{\rm cm}^{-2}{\rm s}^{-1}{\rm sr}^{-1}],
    \label{Eq_e+}
\end{eqnarray}
where $E$ is in units of GeV.
\begin{figure}[ht]
\centering
\includegraphics*[width=2.8in]{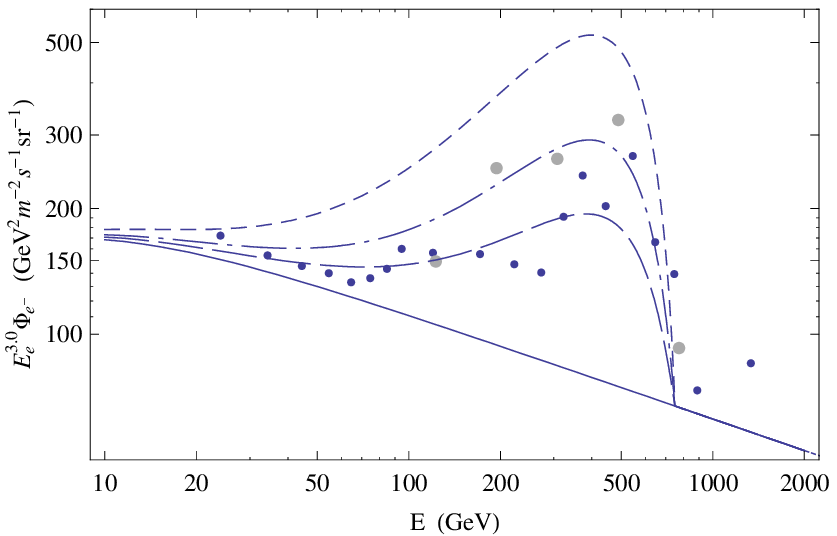}
\includegraphics*[width=2.8in]{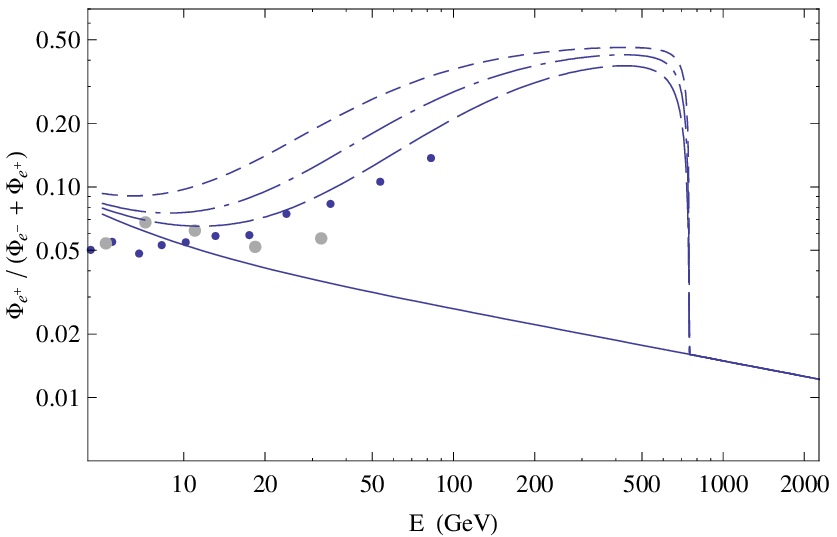}
\caption{Electron energy spectra (left) and   positron fractions
(right) of  the DM decays, where $M_1=10$~GeV, $M_2=1.5$~TeV,
$\tau_2\sim 10^{26}$~s, short-dashed, dot-dashed and long-dashed
lines represent $M=2.5\times10^{15}$, $3\times10^{15}$ and
$3.5\times10^{15}$~GeV, small-black  and large-gray dots stand for
the observations of ATIC and PPB-BETS (left) and PAMELA and HEAT
(right), and solid lines correspond to the backgrounds calculated
from Eqs.~(\ref{Eq_e-})-(\ref{Eq_e+}), respectively.}
 \label{Fig2}
\end{figure}
\begin{figure}[ht]
\centering
\includegraphics*[width=2.8in]{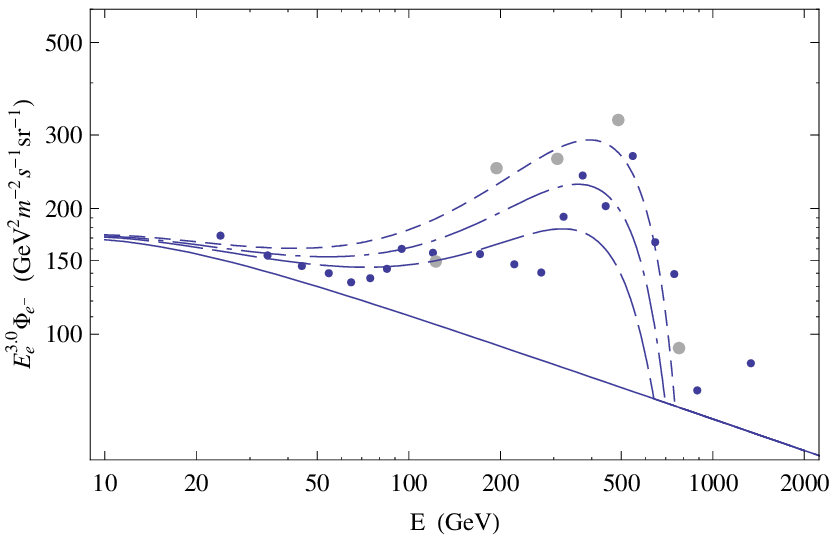}
\includegraphics*[width=2.8in]{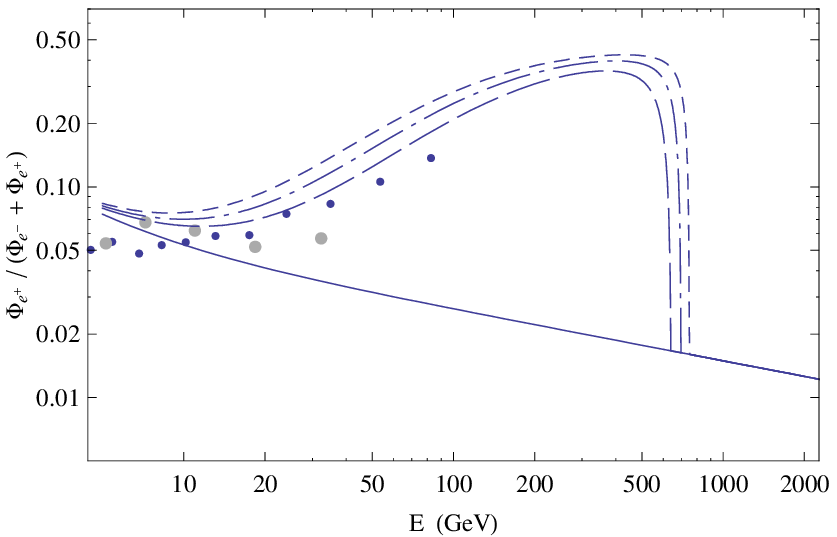}
\caption{Legend is the same as Fig.~\ref{Fig2} but
$M=3\times10^{15}$~GeV, $M_2=2$~TeV and short-dashed, dot-dashed
and long-dashed lines represent $M_1=1$, $1.1$~TeV
 and $1.2$~TeV,
 respectively. }
 \label{Fig3}
\end{figure}
\begin{figure}[ht]
\centering
\includegraphics*[width=2.8in]{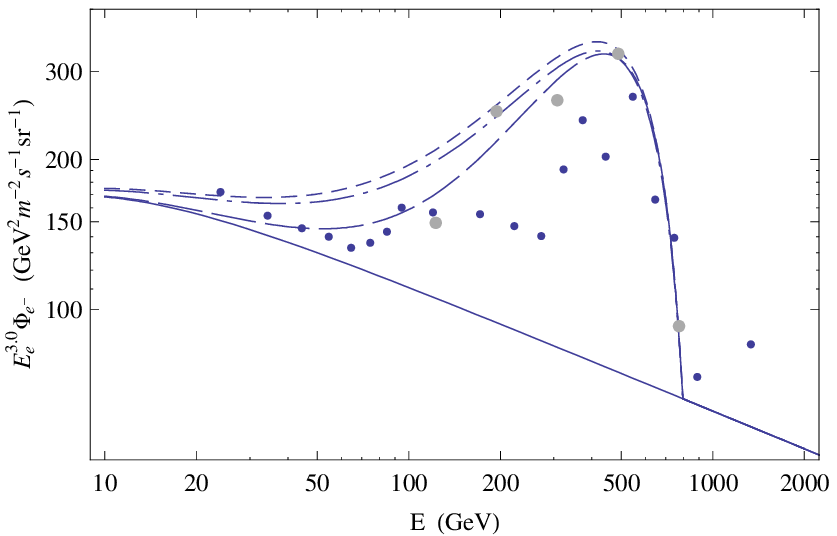}
\includegraphics*[width=2.8in]{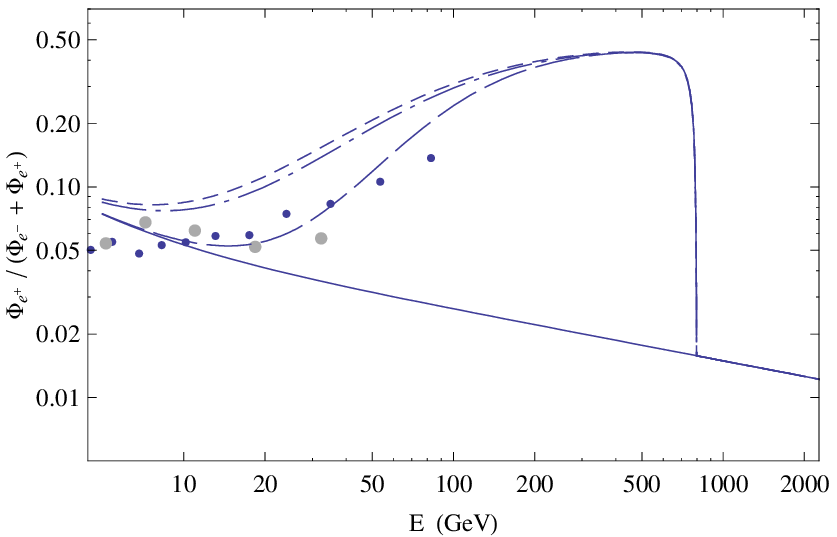}
\caption{ Electron energy spectra (left) and   positron fractions
(right) for the decays of DM particles with $M_2=100$~TeV,
$\tau_2=10^{24}$~s and $M_1=(100-0.8)$~TeV (dashed lines)
 where small-black  and large-gray dots stand for
the observations of ATIC and PPB-BETS (left) and PAMELA and HEAT
(right), and solid,  short-dashed, dot-dashed and long-dashed
lines correspond to the backgrounds, M1, MED and M2,
respectively.}
 \label{Fig4}
\end{figure}
In Fig.~\ref{Fig2}, we show  the electron energy spectra (left)
and  the positron fractions (right)  of  $N_2\to
N_1\ell_i^\pm\ell_j^\mp$ decays for $M_1=10$~GeV, $M_2=1.5$~TeV
and $M=2.5\times10^{15}$~GeV (short-dashed line),
$3\times10^{15}$~GeV (dot-dashed line) and $3.5\times10^{15}$~GeV
(long-dashed line), while Fig.~\ref{Fig3} for
$M=3\times10^{15}$~GeV, $M_2=2$~TeV and $M_1=1$~TeV (short-dashed
line), $1.1$~TeV (dot-dashed line) and $1.2$~TeV (long-dashed
line), respectively. Here, the MED propagation model has been used
and the background is represented by solid lines. In the above
cases, the lifetime of  $N_2$ is of order $10^{26}$~s, while  the
values of $M$ can be achieved by taking $y_{ik}\sim 10^{-3}$ and
$M_\eta\sim10^{12}$~GeV. As seen from Figs.~\ref{Fig2} and
~\ref{Fig3}, the electron energy spectrum is more sensitive to the
parameters in the model in contrast the positron fraction. The P/A
anomaly can be explained by the decays of $N_2$ with the mass
higher than 1.5~TeV. We remark that for a lighter DM particle, the
drop in the electron flux occurs at  a lower energy compared to
the ATIC data. In this sense, our mechanism is complementary to
the annihilations of the DM fermions with masses below 1~TeV,
shown in Refs.~\cite{0901.0176,MaDM}. Our model has more freedom
in DM masses and lifetimes than the models with dominated two-body
decays of DM particles since the drop in the electron flux in our
case is determined by $M_{21}^2/(2M_2)$ rather than (DM mass)/2,
see Refs.~\cite{PAMELAanomaly,Guo_Wu,CMSSM,0901.2168} and
references therein. Hence, the DM lifetime should  be only higher
than the age of the universe ($4.3\times10^{17}$~s). However, for
higher $M_i$, a mass degeneracy is needed. For example, to have
$\tau_2=10^{24}$~s, we need $M_2=10^5$~GeV and
$M_1=(10^5-800)$~GeV. The corresponding electron energy spectra
and positron fractions are shown in Fig.~\ref{Fig4} for the three
propagation models of M1, MED and M2, respectively. We note that
at the energies higher 400~GeV the signals are not sensitive to
the propagation models.
%
%
Production mechanisms of the DM leptons in the early universe and
the upper bounds on their masses from the $\gamma$-rays
observations will be considered elsewhere~\cite{CGZ_in_work}.

Finally, we remark that future collider bounds on the masses of
heavy neutrinos~\cite{heavyNu} are not applicable in this Letter
due to the $Z_2$ symmetry. However, our model can be verified by
precise measurements of the electron spectrum and positron
fraction, since the shapes of the corresponding curves are not
very flexible. In particular, future measurements of the positron
fraction at energies higher than 100~GeV can be crucial in testing
the same origin of the ATIC and PAMELA electron and positron
excesses.

In conclusion, we have investigated a new mechanism to generate
the positron/electron excess from the decays of DM leptons. We
have shown that the observed P/A anomaly can be explained by the
three body decay of the neutral lepton $N_2$ with the mass
$M_2\gtrsim 1.5$~TeV and the lifetime $10^{17}~{\rm
s}\ll\tau_2\lesssim10^{26}$~s. One of the advantages of our
mechanism is that 
there are no requirements for the degeneracy of masses and
unnaturally small couplings or any other enhancement factors.

\section*{Acknowledgements}
 This work is supported in part by
the National Science Council of R.O.C. under Grant Nos: NSC-
97-2112-M-006-001-MY3 and NSC-95-2112-M-007-059-MY3. One of us
(DVZ) would like to thank Prof. Kaoru Hagiwara for the hospitality at KEK and the
exchange program between HEP at Taiwan and KEK.



\begin{thebibliography}{99}

\bibitem{PDG} C. Amsler {\it et al.} [PDG Collaboration], {\it Phys.
Lett.} {\bf B 667} (2008) 1.

\bibitem{PAMELA} O.~Adriani {\it et al.} [PAMELA Collaboration],
Nature {\bf 458} (2009) 607.

\bibitem{ATIC} J.~Chang {\it et al.} [ATIC Collaboration], {\it Nature} {\bf 456} (2008) 362.

\bibitem{PPB-BETS} S.~Torii {\it et al.} [PPB-BETS Collaboration], arXiv:0809.0760 [astro-ph].

\bibitem{HEAT} S.~W.~Barwick {\it et al.} [HEAT Collaboration], {\it Astrophys. J} {\bf 482} (1997) L191.

\bibitem{AMSCollab} M.~Aguilar {\it et al.} [AMS-01 Collaboration], {\it Phys. Lett.} {\bf B 646} (2007) 145.

\bibitem{BoostFactor} J.~Lavalle, Q.~Yuan, D.~Maurin and X.~J.~Bi,
{\it Astron. Astrophys.} {\bf 479} (2008) 427.

\bibitem{Guo_Wu} W.~L.~Guo and Y.~L.~Wu, arXiv:0901.1450 [hep-ph].

\bibitem{PAMELAanomaly} A.~Ibarra and D.~Tran, arXiv:0811.1555 [hep-ph].

\bibitem{CMSSM} I.~Gogoladze, R.~Khalid, Q.~Shafi and H.~Y$\ddot{\rm u}$ksel, arXiv:0901.0923 [hep-ph].

\bibitem{0901.2168} K.~Hamaguchi, F.~Takahashi and T.~T.~Yanagida,
arXiv:0901.2168~[hep-ph]; C.~R.~Chen, F.~Takahashi and
T.~T.~Yanagida, arXiv:0811.0477~[hep-ph].

\bibitem{0812.2075} A.~Arvanitaki {\it et al.}, arXiv:0812.2075
[hep-ph].

\bibitem{Nardi} E.~Nardi, F.~Sannino, A.~Strumia,
arXiv:0811.4153~[hep-ph].

\bibitem{0811.3357} C.~R.~Chen, M.~M.~Nojiri, F.~Takahashi and T.~T.~Yanagida,
arXiv:0811.3357~[astro-ph].

\bibitem{0812.2374} K.~Hamaguchi, S.~Shirai and T.~Yanagida,
arXiv:0812.2374 [hep-ph].

\bibitem{Ma} E.~Ma, {\it Phys. Rev.} {\bf D 73} (2006) 077301.

\bibitem{Gu_Sarkar} P.~Gu and U.~Sarcar, arXiv:0811.0956~[hep-ph].

\bibitem{Ibarra_Tran} A.~Ibarra and D.~Tran, {\it JCAP} {\bf 0807}
(2008) 002.

\bibitem{Chen_Takahashi_Yanagida} C.~R.~Chen, F.~Takahashi and
T.~Yanagida, {\it Phys. Lett.} {\bf B 671} (2009) 71. 

\bibitem{Chen_Takahashi} C.~R.~Chen and F.~Takahashi, {\it JCAP} {\bf 0902} (2009) 004. 

\bibitem{NFW} J.~F.~Navarro, C.~S.~Frenk and S.~D.~M.~White, {\it Astrophysics
J.} {\bf 462} (1996) 563.

\bibitem{MED} T.~Delahaye, R.~Lineros, F.~Donato, N.~Fornengo and P.~Salati, {\it Phys. Rev.} {\bf D 77} (2008) 063527.

\bibitem{BtoC} D.~Maurin, F.~Donato, R.~Taillet and P.~Salati, {\it Astrophys. J.} {\bf 555} (2001) 585.


\bibitem{Blatz} E.~Blatz and J.~Edsjo, {\it Phys. Rev.} {\bf D 59} (1999) 023511.


\bibitem{Moskalenko} I.~Moskalenko and A.~Strong, {\it Astrophys. J.} {\bf 493}
(1998) 694.

\bibitem{0901.0176} X.~J.~Bi, P.~H.~Gu, T.~Li, X.~Zhang,
arXiv:0901.0176[hep-ph].

\bibitem{MaDM} Q.~H.~Cao, E.~Ma and G.~Shaughnessy,
arXiv:0901.1334 [hep-ph].

\bibitem{heavyNu} C.~S.~Chen, C.~Q.~Geng and D.~V.~Zhuridov, {\it Phys. Lett.} {\bf B 666} (2008) 340; A. Ali, A.~V.
Borisov and D.~V. Zhuridov, {\it Phys. Atom. Nucl.} {\bf 68}
(2005) 2061 [{\it Yad. Fiz.} {\bf 68} (2005) 2123]; In: {\it
Particle Physics in Laboratory, Space and Universe} (Singapore,
World Scientific, 2005) P.~66 [hep-ph/0512005];  A.~Ali, A.~V.
Borisov and N.~B. Zamorin, {\it Eur. Phys. J.} {\bf C 21} (2001)
123.

\bibitem{CGZ_in_work} C.~H.~Chen, C.~Q.~Geng and D.~V.~Zhuridov,
in progress.

\end{thebibliography}
\end{document}